# Compactified Reduction from Five to Four Space-Time Dimensions of the Antisymmetric Tensor Field

Michael Kalb[*]

*Physics Department (M-AB2-02), Fairleigh Dickinson University, Madison, New Jersey 07940*
(August 22, 2014)

We employ a Kaluza-Klein dimensional reduction process on the action of the antisymmetric tensor field in five-dimensional space-time. The result is a joint field theory of four-dimensional antisymmetric and vector fields. We write the inhomogeneous Euler-Lagrange equations and homogeneous Bianchi identity equations for the four dimensional field strengths. In these equations the terms that couple the field strengths depend on their variation with the compactified fifth variable. We find that the electric charge current is conserved, but in general the source current of the antisymmetric field is not. The action also displays joint gauge invariance. Observing the fields in a particular Lorentz frame leads to modified Maxwell and antisymmetric field equations. For the Maxwell field the antisymmetric field strengths lead to coupling terms that represent both magnetic charge density and additional induced emf.

## I. INTRODUCTION

String theory has been educated by analogies to the relativistic point particle.[1,2,3] For example, various string actions come about as generalizations of point particle actions and include generalization of relativistic particle invariances. A particularly fruitful generalization stems from the direct inter-particle action formalism explained by Feynman and Wheeler.[4] This formalism leads to an alternative description of charged point particles in an electromagnetic field.[†] Moreover, the formalism leads to the Maxwell field strength equations with source charge currents as well as homogeneous equations due to the definition of the field strengths in terms of a 4-vector potential $a_\mu$. The 4-vector potential obeys inhomogeneous second order partial differential equations and may undergo local gauge transformations without changing the field strengths.

Working by analogy, employing the Feynman-Wheeler method with the Nambu-Gōtō string action[5,6], rather than the point particle action, we may develop an analogous direct inter-string interaction.[7] Instead of leading to a 4-vector potential field theory, the string analogy leads to an antisymmetric tensor potential $A_{\mu\nu} = -A_{\nu\mu}$, which obeys inhomogeneous second order partial differential equations and may undergo local gauge transformations without changing the corresponding field strengths. Interestingly, a combined vector and antisymmetric tensor action[8] supports a Stuckleburg[9] mechanism that can give the antisymmetric tensor field a mass[7,10], much like the Higgs[11] mechanism which gives a vector field a mass in the presence of a scalar field. Both the classical vector and antisymmetric field theories may be extended to any number of space-time dimensions. Moreover, a hierarchy of antisymmetric potentials can be defined in higher space-time dimensions.[12] At each level the Stuckleburg mechanism may be applied, leading to a hierarchy of massive fields.

---

[*] Electronic Address: mikekalb@comcast.net.
[†] In the direct inter-particle theory, all the degrees of freedom are assigned to the charges; the fields are completely dependent on them. We may, however, study the field equations without direct reference to the particles. The formalism also leads to the need for both advanced and retarded potentials. At first sight the advanced potentials appear unphysical. However, it has been shown that advanced potentials may be related to radiation reaction effects.

Field theories in higher dimensions have been explored using the Kaluza-Klein[13] method. For example, a 5 space-time dimensional gravitational field may be shown to be equivalent to a combined 4 space-time dimensional gravitational field, vector field and scalar field. This process of reduction from 5 to 4 space-time dimensions is accomplished by singling out a particular direction in space, thereby breaking the 5-dimensional space-time symmetry to a residual 4-dimensional space-time symmetry.

We here apply the Kaluza-Klein method to the action of a 5-dimensional space-time (5D) antisymmetric field. The action is invariant under the Lorentz group of space-time transformations. By singling out a particular spacial direction, we break the full 5D symmetry down to a 4D symmetric theory. The result is a combined interacting antisymmetric tensor and vector field theory in 4D. Moreover, by choosing a particular time direction in this resulting Minkowski manifold, we produce a set of interacting field strengths in an SO(3) symmetric theory. One may then identify a subset of these fields with the Maxwell fields, while the remaining field strengths refer to the 4D antisymmetric tensor field. The fields are coupled due to residual dependence on the singled-out spacial dimension. Nevertheless, vanishingly small field variation with respect to this coordinate leads to a decoupling of the Maxwell fields from the antisymmetric tensor fields. We may assume that the singled-out spacial dimension is compactified with periodic boundary conditions.

### A. Five Dimensional Action

In five space-time dimensions, we postulate the Lorentz invariant action for the interacting antisymmetric field theory as

$$S = \int d^5 x \mathcal{L}$$
$$\mathcal{L} = \tfrac{1}{12} \mathcal{F}^{ABC} \mathcal{F}_{ABC} + \tfrac{1}{2} \mathcal{J}^{AB} \mathcal{A}_{BC}.$$
(1.1)

Capital Roman indices enumerate the 5D space-time dimensions (i.e., $A = 0,1,2,3,4.$). Here $\mathcal{L}$ is the Lagrangian density; $\mathcal{A}_{AB} = -\mathcal{A}_{BA}$ is the antisymmetric potential; the completely antisymmetric field strength is $\mathcal{F}_{ABC} \equiv \partial_A \mathcal{A}_{BC} + \partial_B \mathcal{A}_{CA} + \partial_C \mathcal{A}_{AB}$, and $\mathcal{J}_{AB}$ is an external antisymmetric current that directly couples to the potential.

We may organize the indices in 5-space-time according to

$$\begin{aligned} A &= (\mu, 4) \\ \mu &= (0, i) \\ i &= (1,2,3). \end{aligned}$$
(1.2)

Thus, Greek indices correspond to 4D space-time, and lower case Roman indices enumerate the remaining 3D spacial dimensions. Our conventions for the Minkowski metrics are:

$$\eta_{AB} = \begin{cases} 1 & A = B = 0 \\ -1 & A = B = 1,2,3,4 \\ 0 & \text{otherwise} \end{cases} \qquad \eta_{\mu\nu} = \begin{cases} 1 & \mu = \nu = 0 \\ -1 & \mu = \nu = 1,2,3 \\ 0 & \text{otherwise} \end{cases}. \tag{1.3}$$

We also define

$$\varepsilon_{0123} \equiv 1 \equiv \varepsilon_{123}, \tag{1.4}$$

for the Levi-Civita densities of the subspaces. Finally, we measure velocities in terms of the speed of light, and thus $c = 1$.

From the antisymmetry of the 5D potentials, we may deduce that

$$\mathscr{F}^{ABC} = \mathscr{F}^{BCA} = \mathscr{F}^{CAB} = -\mathscr{F}^{CBA} = -\mathscr{F}^{BAC} = -\mathscr{F}^{ACB}$$
$$\mathscr{G}^{BC} = -\mathscr{G}^{CB}. \tag{1.5}$$

We also define the following special directional derivatives of any function of 5D space-time, $w(x^A)$.

$$w'(x^A) \equiv \partial_4 w(x^A)$$
$$\dot{w}(x^A) \equiv \partial_0 w(x^A). \tag{1.6}$$

### B.  Euler-Lagrange Equations and Bianchi Identities

The inhomogeneous Euler-Lagrange equations obtained by extremizing the 5D action over variations of the antisymmetric potentials $\mathscr{A}_{AB}$, are

$$\mathscr{G}^{BC} = \partial_A \mathscr{F}^{ABC}. \tag{1.7}$$

Since the potential fields $\mathscr{A}_{AB}$ are antisymmetric, the Bianchi identities also apply as constraints on the field strengths:

$$\partial_A \mathscr{F}_{BCD} + \partial_B \mathscr{F}_{CDA} + \partial_C \mathscr{F}_{DAB} + \partial_D \mathscr{F}_{ABC} = 0. \tag{1.8}$$

These are the antisymmetric tensor homogeneous equations.

## II. DIMENSIONAL REDUCTION

Let us choose a 5D frame where spacial dimension $x^4$ is assumed to be compactified with radius $R$. Perform the dimensional reduction on the inhomogeneous equations by separating the spacial dimension described by $x^4$. We obtain.

$$\mathscr{J}^{BC} = \partial_\mu \mathscr{F}^{\mu BC} + \mathscr{F}'^{4BC}. \tag{2.1}$$

Thus,

$$\begin{aligned}\mathscr{J}^{\nu C} &= \partial_\mu \mathscr{F}^{\mu\nu C} + \mathscr{F}'^{4\nu C} \\ \mathscr{J}^{4C} &= \partial_\mu \mathscr{F}^{\mu 4C} + \mathscr{F}'^{44C} = \partial_\mu \mathscr{F}^{\mu 4C}.\end{aligned} \tag{2.2}$$

These imply that

$$\begin{aligned}\mathscr{J}^{\nu\rho} &= \partial_\mu \mathscr{F}^{\mu\nu\rho} + \mathscr{F}'^{\nu\rho 4} \\ \mathscr{J}^{\nu 4} &= \partial_\mu \mathscr{F}^{\mu\nu 4}.\end{aligned} \tag{2.3}$$

We have used the antisymmetry of $\mathscr{J}^{AB}$ and $\mathscr{F}^{ABC}$; hence $\mathscr{J}^{44} = 0$ and $\mathscr{F}^{44C} = 0$.

If we define

$$\begin{aligned}J^{\nu\rho} &\equiv \mathscr{J}^{\nu\rho} & F^{\mu\nu\rho} &\equiv \mathscr{F}^{\mu\nu\rho} \\ j^\nu &\equiv \mathscr{J}^{\nu 4} & f^{\mu\nu} &\equiv \mathscr{F}^{\mu\nu 4},\end{aligned} \tag{2.4}$$

then equations (2.3) become

$$\partial_\mu F^{\mu\nu\rho} + f'^{\nu\rho} = J^{\nu\rho} \tag{2.5a}$$

$$\partial_\mu f^{\mu\nu} = j^\nu. \tag{2.5b}$$

Other than the derivative of $f^{\nu\rho}$ with respect to the compactified variable, the first equation is the Euler-Lagrange equation of motion of a 4D antisymmetric tensor field strength produced by an antisymmetric source current $J^{\nu\rho}$. This current is approximately conserved if $\partial_\nu f'^{\nu\rho} \approx 0$. The second set of equations has the form of the inhomogeneous equations of motion of a Maxwell field strength, with a conserved vector source (electric charge) current $j^\nu$.

The inhomogeneous 5D Bianchi identities may be treated in a similar manner to yield the set of complementary 4D homogeneous equations:

$$0 = \varepsilon^{\mu\nu\rho\sigma} \partial_\mu F_{\nu\rho\sigma} \tag{2.6a}$$

$$0 = \partial_\mu f_{\nu\rho} + \partial_\nu f_{\rho\mu} + \partial_\rho f_{\mu\nu} - F'_{\mu\nu\rho}. \tag{2.6b}$$

The first equation is the usual Bianchi identity satisfied by a 4D antisymmetric tensor field strength. Except for $F'_{\mu\nu\rho}$, the second equations are the Bianchi identities satisfied by a 4D Maxwell field strength.

We see therefore that the 5D action (1.1) of an antisymmetric field produces the equations of motion of a 4D antisymmetric field plus the equations of a 4D vector field in the limit of small field variation in the compactified dimension. On the other hand, if this field variation is significant, the field strengths become coupled. The dimensional reduction proceeds similarly for higher dimensions than five.

### A. Current Conservation

The 5D field strength is antisymmetric under index exchange. Therefore, the 5D current is conserved.

$$\partial_B \partial_A \mathcal{F}^{ABC} = 0 = \partial_B \mathcal{J}^{BC}. \tag{2.7}$$

Separating out in the $x^4$ direction, equation (2.7) leads to

$$\begin{aligned} 0 &= \partial_\mu J^{\mu\nu} + j'^\nu \\ 0 &= \partial_\mu j^\mu. \end{aligned} \tag{2.8}$$

The 4D tensor current source is conserved only if the variation of the vector current with the compactified variable is small, (i.e., $j'^\nu \approx 0$). The vector (electric charge) current source is conserved.

### B. Potential Field Equations

Starting with the antisymmetric 5D inhomogeneous Euler-Lagrange equations (1.7)

$$\mathcal{J}^{BC} = \partial_A \mathcal{F}^{ABC}, \tag{1.7}$$

and the 5D field strength expressed in terms of the antisymmetric potentials, we find the (second order) equations of motion of the potentials

$$\begin{aligned} J^{\nu\rho} &= \partial_\mu \left( \partial^\mu A^{\nu\rho} + \partial^\nu A^{\rho\mu} + \partial^\rho A^{\mu\nu} \right) + \left( A''^{\nu\rho} + \partial^\nu a'^\rho - \partial^\rho a'^\nu \right) \\ j^\nu &= \partial_\mu \left( \partial^\mu a^\nu - \partial^\nu a^\mu \right) + \partial_\mu A'^{\mu\nu}. \end{aligned} \tag{2.9}$$

Here we have defined

$$A^{\mu\nu} \equiv \mathcal{Q}^{\mu\nu}$$
$$a^{\mu} \equiv \mathcal{Q}^{\mu 4}. \tag{2.10}$$

Thus, the 4D inhomogeneous equations for the tensor and vector potentials have their usual forms if the potentials vary slowly with the compactified variable. Otherwise, the 4D potential equations are coupled.

The 5D Bianchi identities produce the homogeneous 4D field strength equations (2.6)

$$0 = \partial_\mu F_{\nu\rho\sigma} + \partial_\nu F_{\rho\sigma\mu} + \partial_\rho F_{\sigma\mu\nu} + \partial_\sigma F_{\mu\nu\rho}$$
$$0 = \partial_\mu f_{\nu\rho} + \partial_\nu f_{\rho\mu} + \partial_\rho f_{\mu\nu} - F'_{\mu\nu\rho}, \tag{2.11}$$

which are solved by

$$F_{\mu\nu\rho} = \partial_\mu A_{\nu\rho} + \partial_\nu A_{\rho\mu} + \partial_\rho A_{\mu\nu}$$
$$f_{\mu\nu} = \partial_\mu a_\nu - \partial_\nu a_\mu + A'_{\mu\nu}. \tag{2.12}$$

Again, the 4D homogeneous equations for the tensor and vector field strengths have their usual forms if the tensor field strength varies slowly with the compactified variable. Otherwise, the 4D field strength homogeneous equations are coupled. Note that generally the Maxwell field strength $f_{\mu\nu}$ has the added term, $A'_{\mu\nu}$, which modifies the usual potential solutions of the homogeneous equations. Thus,

$$f_{\mu\nu} \neq \partial_\mu a_\nu - \partial_\nu a_\mu, \tag{2.13}$$

where $a_\mu$ is the usual vector potential,[14] if $A'_{\mu\nu} \approx 0$.*

### C. Gauge Invariance

The 5D action for the antisymmetric field displays gauge invariance. If we let

$$\mathcal{Q}_{AB} \to \mathcal{Q}_{AB} + \partial_A \mathcal{G}_B - \partial_B \mathcal{G}_A, \tag{2.14}$$

where $\mathcal{G}_A$, the gauge parameter, is a differentiable 5-vector function of 5D space-time, then the 5D antisymmetric field strength $\mathcal{F}_{ABC}$ is invariant (and therefore so is the action).† This local

---

* As we shall see, if the Maxwell field strength obeys (2.13), then magnetic charges are predicted. See reference (14) for another example.
† Interestingly, one may always add a 5-gradient $\partial_A \mathcal{X}$ to $\mathcal{G}_A$ without changing the transformed potential. In Maxwell theory this is analogous to adding a global constant to the gauge parameter.

gauge transformation leads to the following local gauge transformations on the reduced dimensional potentials

$$A_{\mu\nu} \to A_{\mu\nu} + \partial_\mu \Lambda_\nu - \partial_\nu \Lambda_\mu$$
$$a_\mu \to a_\mu + \partial_\mu \lambda - \Lambda'_\mu , \quad (2.15)$$

with

$$\Lambda_\mu \equiv \mathcal{G}_\mu$$
$$\lambda \equiv \mathcal{G}_4 . \quad (2.16)$$

Note that the 4D tensor field replicates the 5D gauge transformation, but the 4D vector field has an additional term that couples its gauge transformation to that of the tensor, forming a joint gauge transformation. The additional term is small if the 4-vector gauge function has small variation with the compactified variable, and then the usual gauge transformation remains.

### D. Lagrangian

The Lagrangian may be put in a reduced dimensional form by repeating the above process. We then obtain

$$\mathcal{L} = \tfrac{1}{12} \mathcal{F}^{ABC} \mathcal{F}_{ABC} + \tfrac{1}{2} \mathcal{J}^{AB} \mathcal{Q}_{AB}$$
$$= \tfrac{1}{12} F^{\mu\nu\rho} F_{\mu\nu\rho} + \tfrac{1}{4} f^{\mu\nu} f_{\mu\nu} + \tfrac{1}{2} J^{\mu\nu} A_{\mu\nu} + j^\mu a_\mu . \quad (2.17)$$

Keep in mind, however, that $f_{\mu\nu} = \partial_\mu a_\nu - \partial_\nu a_\mu + A'_{\mu\nu}$. Thus, unlike $j^\mu$, $J^{\mu\nu}$ is not conserved in 4D.

### III. SPECIFIC LORENTZ FRAME

#### A. Specify 4D Lorentz Frame

Let us observe the system of fields in a particular Lorentz frame where

$$t = x_0$$
$$\vec{x} = (x, y, z) = (x_1, x_2, x_3) . \quad (3.1)$$

Equation (2.5b) may then be rewritten as

$$\vec{\nabla} \cdot \vec{E} = \rho$$
$$\vec{\nabla} \times \vec{B} - \dot{\vec{E}} = \vec{j} . \quad (3.2)$$

While equation (2.6b) produces

$$\vec{\nabla} \cdot \vec{B} = K'$$
$$\vec{\nabla} \times \vec{E} + \dot{\vec{B}} = \vec{R}',$$
(3.3)

where

$$\begin{aligned}
(\vec{E})_i &= f_{i0} & \rho &= -j_0 \\
(\vec{B})_i &= \tfrac{1}{2}\varepsilon_{ijk} f_{jk} & (\vec{j})_i &= j_i \\
(\vec{R})_i &= \tfrac{1}{2}\varepsilon_{ijk} F_{jk} \\
K &= \tfrac{1}{3}\varepsilon_{ijk} F_{ijk}.
\end{aligned}$$
(3.4)

These modified forms of Maxwell's equations contain contributions from components of the tensor field. Note the presence of magnetic charge density $K'$ and the additional source of induced emf $\vec{R}'$.

In the same Lorentz frame the tensor field strength equations become

$$\vec{\nabla} \times \vec{R} + \vec{E}' = \vec{\sigma}$$
$$\dot{\vec{R}} - \vec{\nabla} K + \vec{B}' = \vec{\tau},$$
(3.5)

where

$$\begin{aligned}
(\vec{\tau})_i &= \tfrac{1}{2}\varepsilon_{ijk} J_{jk} \\
(\vec{\sigma})_i &= J_{i0}.
\end{aligned}$$
(3.6)

In addition the Bianchi identity, equation (2.6a), gives

$$\vec{\nabla} \cdot \vec{R} - \dot{K} = 0.$$
(3.7)

If there are no tensor currents $(\vec{\tau}, \vec{\sigma})$, and if the variation of $(\vec{E}, \vec{B})$ with the compactified variable is small, then the solutions to equations (3.5) and (3.7) are

$$K \approx \dot{\phi}$$
$$\vec{R} \approx \vec{\nabla}\phi,$$
(3.8)

Where $\phi$ approximately obeys

$$\ddot{\phi} - \nabla^2 \phi \approx 0.$$
(3.9)

Under these conditions the tensor field can be approximately represented by a massless scalar potential. However, if there is significant coupling to the Maxwell field strengths and/or a tensor source current, then equations (3.8) are no longer solutions to equation (3.7).

Since the 5D tensor source current is conserved, it leads to relations among the 4D tensor and vector source currents. We find that

$$
\begin{aligned}
0 &= \dot{\rho} + \vec{\nabla} \cdot \vec{j} \\
0 &= -\vec{\nabla} \cdot \vec{\sigma} + \rho' \\
0 &= \dot{\vec{\tau}} + \vec{j}' \\
0 &= \dot{\vec{\sigma}} - \vec{\nabla} \times \vec{\tau} .
\end{aligned}
\tag{3.10}
$$

As already noted the vector (charge) current is conserved. However, the $\vec{\tau}$ and $\vec{\sigma}$ currents depend on the variation of the charge density and charge 3-current density with the compactified coordinate $x^4$.

## IV. CONCLUSION

We found that an antisymmetric tensor field in 5D Minkowski space-time can be recast as an antisymmetric tensor field in 4D Minkowski space-time interacting with a 4D Maxwell field. The extent of the interaction depends on the variation of the field strengths with the singled-out compactified coordinate. Coupling of the 4D fields occurs in both the inhomogeneous Euler-Lagrange equations and the Bianchi identity constraints. As a result the Maxwell field theory contains a source of magnetic charge and an additional source of emf. These come from derivatives of components of the antisymmetric field strength with respect to the compactified variable. The antisymmetric field strengths have additional sources from derivatives of Maxwell field strengths with respect to the compactified variable. Since the 5D action is invariant under the usual gauge transformations of antisymmetric potentials, the 4D antisymmetric and vector potentials undergo joint gauge transformations. The 4D antisymmetric potentials inherit their usual gauge transformations, while the Maxwell potentials have a modified gauge transformation which depends on the derivative with respect to the compactified variable of the local gauge parameter.

Although the Maxwell current source continues to be conserved in 4D, the antisymmetric current source is generally not. As expected, the "leakage" depends on Maxwell current source derivatives with respect to the compactified variable.

The 5D equations are consistent, and thus the interacting field theory between the antisymmetric tensor field and vector field is as well. The Kaluza-Klein process applied to this 5D field theory represents another example of how antisymmetric tensor fields and vector fields appear to have simple complementary relationships to each other.

## ACKNOWLEDGEMENTS

The author thanks the Fairleigh Dickinson University Department of Mathematics, Computer Science and Physics at Madison for its support.